\documentclass[10pt]{iopart}
\usepackage{epsfig,graphicx,mathrsfs}
\usepackage{amssymb,tikz,wrapfig}
\usepackage{txfonts}
\usepackage[titletoc]{appendix} 
\usepackage{colordvi}

\graphicspath{{graphics/}}
\newcommand{\N}{{\rm I\!N}}

\newcommand{\vsp}{\vspace*{3mm}}

\newcommand{\order}{{\cal O}}

\newcommand{\ba}{\mbox{\boldmath $a$}}
\newcommand{\bb}{\mbox{\boldmath $b$}}
\newcommand{\bc}{\mbox{\boldmath $c$}}

\newcommand{\bk}{\mbox{\boldmath $k$}}

\newcommand{\bm}{\mbox{\boldmath $m$}}
\newcommand{\bn}{\mbox{\boldmath $n$}}
\newcommand{\bq}{\mbox{\boldmath $q$}}

\newcommand{\bX}{\mbox{\boldmath $X$}}

\newcommand{\bbeta}{\mbox{\boldmath $\beta$}}

\newcommand{\btheta}{\mbox{\boldmath $\theta$}}

\newcommand{\bphi}{\mbox{\boldmath $\phi$}}

\newcommand{\btau}{\mbox{\boldmath $\tau$}}

\newcommand{\bbox}{{{\bullet}}}
\newcommand{\here}{\makebox(0,0)}

\begin{document}
\title[]{Entropies of tailored random graph ensembles:  bipartite graphs, generalised degrees,  and node neighbourhoods}
\author{ES Roberts$^{\dag \ddag}$ and ACC Coolen$^{\dag \S}$} 

\address{
${\dag}~$Institute for Mathematical and Molecular Biomedicine, King's College London,  Hodgkin Building,
London SE1 1UL, United Kingdom}
\address{
${\ddag~}$Randall Division of Cell and Molecular Biophysics, King's College London, New
Hunts House, London SE1 1UL, United Kingdom}
\address{$\S~$
London Institute for Mathematical Sciences, 35a South St, Mayfair, London W1K 2XF, United Kingdom}

\pacs{89.70.Cf, 89.75.Fb, 64.60.aq}

\ead{ekaterina.roberts@kcl.ac.uk ton.coolen@kcl.ac.uk}

\begin{abstract}
We calculate explicit formulae for the  Shannon entropies  of several families of tailored random graph ensembles for which no such formulae were as yet available,  in leading orders in the system size.  These include bipartite graph ensembles with imposed (and possibly distinct) degree distributions for the two node sets, graph ensembles constrained by specified node neighbourhood distributions, and graph ensembles constrained by specified generalised degree distributions. 
\end{abstract}

\section{Introduction}

Networks are powerful and popular tools for characterising large and complex interacting particle systems. They have become extremely valuable in physics, biology, computer science, economics, and the social sciences. One approach is to 
 quantify the implications of having topological patterns in networks and graphs, by viewing these patterns as constraints on a random graph ensemble. 
This provides a way to measure and compare topological features from the rational point of view of whether they are present in a large or small number of possible networks. Precise definitions of random graph ensembles with controlled topological characteristics also allow us to generate systematically graphs and networks which are tailored to have features in common with those observed in a given application domain, either for the purpose of statistical mechanical process modelling or to serve as `null models' against which to test the importance of observations in real-world networks.  

A previous paper \cite{Annibale09} considered tailored random graph ensembles with controlled degree distribution and degree-degree correlations; the more recent \cite{Roberts11} covered the case of directed networks. In each case, the strategy is to calculate the Shannon entropy, from which we can deduce the effective number of graphs in the ensemble. Related quantities such as complexity of typical graphs from the ensemble and information-theoretic distances between graphs naturally follow from the entropy, or can be calculated using similar methods. 

In this paper we calculate, in leading order, the Shannon entropies of three as yet unsolved families of random graph ensembles, constrained by three different conditions: a bipartite constraint with imposed degree distributions in the two nodes sets,  a neighbourhood distribution (where the neighbourhood of a node is defined as its own degree, plus the  degree values of the nodes connected to it), and an imposed generalised degree distribution. These are each interesting in their own right as stand-alone results, and turn out to be closely linked. The first two cases can be resolved exactly, and give practical analytical expressions. The generalised degree case was already partially studied in \cite{bianconi2008entropies}, with only limited success, and here we require a plausible but as yet unproven conjecture to find an explicit formula for the entropy. 

 The generalised degrees concept appears in the literature in various forms. For example, the authors of \cite{Faudre}, measured the number of direct neighbours $s$ of a subset of $t$ nodes. They derive conditions based on their definition of general degrees which can ensure that (for some given $m$ and $d$ ) there are at least $m$ internally disjoint paths of length at most $d$. The diameter of the network is an obvious corollary - the smallest $d$ corresponding to $m \geq 1$. These results can be applied to questions of robustness of networks.  The authors of \cite{rogers2010spectral} studied the spectral density of random graphs with hierarchically constrained topologies. This includes consideration of generalised degrees, as well as more general community structures. Using the replica method, in a similar way to \cite{bianconi2008entropies}, they achieve a form analogous to equation \eref{eq:intermediate_form}. They proceed numerically from that point, hence our approach to an analytical solution presented in equation \eref{eq:final_answer} is entirely novel.

\section{Definitions and notation}
\label{sec:gd_notation}

 We consider ensembles of directed and nondirected random graphs.
Each graph is defined by its adjacency matrix $\bc=\{c_{ij}\}$, with $i,j\in\{1,\ldots,N\}$ and
with $c_{ij}\in\{0,1\}$ for all $(i,j)$. Two nodes $i$ and $j$ are connected by a directed link $j\to i$ if and only if  $c_{ij}=1$. We put $c_{ii}=0$ for all $i$. In nondirected graphs one has $c_{ij}=c_{ji}$ for all $(i,j)$, so $\bc$ is symmetric. The degree of a node $i$ in a nondirected graph is the number of its neighbours, $k_i = \sum_j c_{ij}$.  In directed graphs we distinguish between in- and out-degrees,  $k_i^{\rm{in}} = \sum_j c_{ij}$ and $k_i^{\rm{out}} = \sum_j c_{ji}$. They count the number of in- and out-bound links at a node $i$. A bipartite graph is one where the nodes can be divided into two disjoint sets, such that $c_{ij}=0$ for all $i$ and $j$ that belong to the same set.

\begin{figure}[t]
\hspace*{18mm}\unitlength=0.29mm
\begin{picture}(200,190)

\put(0,0){\includegraphics[width=198\unitlength]{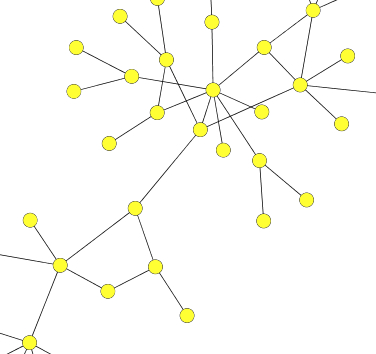}}

\put(105.8,117.3){\here{\LARGE$\bbox$}}
\put(89,110){$i$}

\put(240,150){$k_i=4$}
\put(240,130){$(k_i,m_i)=(4,20)$}
\put(240,110){$n_i=(k_i;~\{\xi_i^s\})=(4;~3,4,6,7)$}

\end{picture}\vspace*{-0mm}

\caption{Illustration of our definitions of local topological chacteristics in non-directed graphs. At the minimal level one specifies for each node $i$ (black vertex in the picture) only the degree $k_i=|\partial_i|=\sum_j c_{ij}$  (the number of its neighbours). At the next level of detail one provides for each node the generalised degree $(k_i,m_i)$, in which $m_i=\sum_{j\in\partial_i}k_j=\sum_{j}c_{ij}k_j$ is the number of length-two paths starting in $i$. This is then generalised to include the actual degrees in the set  $\partial_i$, by giving $n_i=(k_i;~\{\xi_i^s\})$ (the `local neighbourhood'), in which the $k_i$ integers $\{\xi_i^s\}$ give the degrees of the nodes connected to $i$. To avoid ambiguities we adopt the ranking convention $\xi_i^1\leq\xi_i^2 \leq \ldots\leq \xi_i^{k_i}$. Note that $m_i=\sum_{j\in\partial_i}k_j=\sum_{s=1}^{k_i}\xi_i^s$.}
\label{fig:neighbourhood}
\end{figure}

We define the  set of neighbours of a node $i$ in a nondirected graph as $\partial_i=\{j|~c_{ij}=1\}$. 
Hence $k_i=|\partial_i|$. To characterise a graph's topology near $i$ in more detail we can define the generalised degree of $i$ as the pair $(k_i,m_i)$, where $m_i=\sum_j c_{ij}k_j$ counts the number of length-two paths starting in $i$. 
The concept of a generalised degree is discussed in  \cite{Newmanbook}. 
Even more information is contained in the {\em local neighbourhood}
\begin{eqnarray}
 n_i&=&(k_i;~\{\xi_i^s\}), 
\label{eq:neighbourhood}
\end{eqnarray} 
in which the ordered integers $\{\xi_i^s\}$ give the  degrees of the $k_i$ neighbours $j\in\partial_i$. See also Fig. \ref{fig:neighbourhood}.  Since $m_i=\sum_{s\leq k_i}\xi_i^s$, the neighbourhood $n_i$ provides more granular information that complements that in the generalised degree $(k_i,m_i)$. 
We will use 
bold symbols when local topological parameters are defined for every node in a network, e.g. $\bk = (k_1,\ldots, k_N)$ and $\bn=((k_1;~\{\xi_1^s\}),\ldots,(k_N;~\{\xi_N^s\}))$. 
Generalisation to directed graphs is straightforward. Here $\partial_i=\{j|~c_{ij}\!+\!c_{ji}>0\}$, and  the local neighbourhood would be defined as $n_i=(\vec{k}_i;~\{\vec{\xi}_i^s\})$ with the $k_i$ pairs $\vec{\xi}_i^s=(k^{s,\rm in},k^{s, \rm out})$ now giving both the in- and out-degrees of the neighbours of $i$.
\vsp

Our tailored random graph ensembles will  be of the following form, involving $N$ built-in local (site specific) topological constraints of the type discussed above, which we will for now write generically as $X_i(\bc)$, and with the usual abbreviation $\delta_{\ba,\bb}=\prod_i \delta_{a_i,b_i}$:
\begin{eqnarray}
\label{eq:ensemble}
p(\bc)&=& \sum_{\bX} p(\bX) \, p(\bc|\bX)~~~~~~p(\bX)=\prod_{i} p(X_i)
\\ 
p(\bc|\bX)&=& Z^{-1}(\bX)  \delta_{\bX,\bX(\bc)},~~~~~~
Z(\bX)=\sum_{\bc}\delta_{\bX,\bX(\bc)}
\end{eqnarray}
The values $X_i$ for the local features are for each $i$ drawn randomly and independently from $p(X)$, after which one generates a graph $\bc$ randomly and with uniform probabilities from the set of graphs that satisfy the $N$ demands $X_i(\bc)=X_i$. 
The empirical distribution $p(X|\bc)=N^{-1}\sum_i\delta_{X,X_i(\bc)}$ of local features will be random, but the law of large numbers ensures that for $N\to\infty$ it will converge to the chosen $p(X)$ in (\ref{eq:ensemble})  for any graph realisation, and the above definitions guarantee that its ensemble average will be identical to $p(X)$ for any $N$, 
\begin{eqnarray}
\sum_{\bc}p(\bc)p(X|\bc)&=&\frac{1}{N}\sum_{i}\sum_{\bX}p(\bX) \sum_{\bc}\frac{\delta_{\bX,\bX(\bc)}}{Z(\bX)}\delta_{X,X_i(\bc)}
=p(X)
\end{eqnarray}
If we aim to impose upon our graphs only a degree distribution we choose $X_i(\bc)=k_i(\bc)$.  Building in a distribution of generalised degrees  corresponds to $X_i(\bc)=(k_i(\bc),m_i(\bc))$.  If we seek to prescribe the distribution of all local neighbourhoods (\ref{eq:neighbourhood}) we choose $X_i(\bc)=n_i(\bc)$.

A further quantity which will play a role in subsequent calculations is the joint degree distribution of connected nodes. For nondirected graphs it is defined as
\begin{eqnarray}
W(k, k^\prime|\bc) &=& \frac{\sum_{ij} c_{ij} \delta_{k, k_i} 
\delta_{k^\prime, k_j}
}
{
\sum_{ij} c_{ij}
}
\label{eq:Wempirical}
\end{eqnarray}
and its average over the ensemble (\ref{eq:ensemble}) is given by
\begin{eqnarray}
W(k, k^\prime) &=& \sum_{\bX} p(\bX) \sum_{\bc}
W(k, k^\prime| \bc)\frac{\delta_{\bX,\bX(\bc)}}{Z(\bX)}
\label{eq:W}
\end{eqnarray}

\vsp

In this paper we study the leading orders in the system size $N$ of the Shannon entropy per node of the above tailored random graph ensembles (\ref{eq:ensemble}), from which the effective number of graphs with the prescribed distribution $p(X)$ of features follows as ${\cal N}=\exp(NS)$:
\begin{eqnarray}
S&=&-\frac{1}{N}\sum_{\bc}p(\bc)\log p(\bc)
\nonumber
\\
&=& -\frac{1}{N} \sum_{\bX} \frac{\prod_{i} p(X_i)}{Z(\bX)}  \sum_{\bc} \delta_{\bX,\bX(\bc)}\log\Big[
\sum_{\bX^\prime} \frac{\prod_{j} p(X_j^\prime)}{Z(\bX^\prime) } \delta_{\bX^\prime,\bX(\bc)}\Big]
\nonumber
\\
&=&  -\frac{1}{N}\sum_{\bX} \frac{\prod_{i} p(X_i)}{Z(\bX)}  \sum_{\bc} \delta_{\bX,\bX(\bc)}\log\Big[
\frac{\prod_{j} p(X_j)}{Z(\bX) }\Big]
\nonumber
\\
&=& 
\sum_{\bX} p(\bX) S(\bX) 
-\sum_{X} p(X)\log p(X)
\label{eq:S}
\end{eqnarray}
with
\begin{eqnarray}
S(\bX)&=& 
\frac{1}{N}\log Z(\bX)=\frac{1}{N}\log \sum_{\bc}\delta_{\bX,\bX(\bc)}
\label{eq:core}
\end{eqnarray}
The core of the entropy calculation is determining the leading orders in $N$ of $S(\bX)$, which is the Shannon entropy per node of the ensemble $p(\bc|\bX)$ in which all node-specific  values $\bX=(X_1,\ldots,X_N)$ are constrained. 
For $p(X)=p(k)$ this calculation has already been done in \cite{Annibale09,Roberts11}. 
For $p(X)=p(k,m)$ it has only partly been done \cite{bianconi2008entropies}. 
Here we investigate the relation between the entropies of the $p(k)$ and $p(k,m)$ ensembles  and the entropy of the ensemble in which the distribution $p(n)$ of local neighbourhoods (\ref{eq:neighbourhood}) is imposed. 


\section{Building blocks of the entropy calculations}

\subsection{Relations between feature distributions for nondirected graphs}

Since the generalised degrees $(k_i,m_i)$ can be calculated from the local neighbourhoods (\ref{eq:neighbourhood}) for any graph $\bc$, it is clear that the empirical distribution $p(k,m|\bc)=N^{-1}\sum_{i} \delta_{k,k_i(\bc)}\delta_{m,m_i(\bc)}$ for any graph can be calculated from the empirical neighbourhood distribution $p(n|\bc)=N^{-1}\sum_i\delta_{n,n_i(\bc)}$. If we denote with $k(n)$ the central degree $k$ in $n=(k;~\{\xi^s\})$, we indeed  obtain
\begin{eqnarray}
p(k,m|\bc)&=& \frac{1}{N}\sum_{i} \delta_{k,k_i(\bc)}\delta_{m,m_i(\bc)}\sum_{n}\delta_{n,n_i}
=\sum_{n} p(n)~\delta_{k,k(n)}\delta_{m,\sum_{s\leq k(n)}\xi^s}
\end{eqnarray}
Less trivial is the statement that also the distribution $W(k,k^\prime|\bc)$ of 
(\ref{eq:Wempirical}) can be written in terms of $p(n|\bc)$. Using $\sum_{ij}c_{ij}=N\bar{k}(\bc)$, with $\bar{k}(\bc)=N^{-1}\sum_i k_i(\bc)$ we obtain
\begin{eqnarray}
W(k,k^\prime|\bc)&=& \frac{\sum_{i}\delta_{k,k_i(\bc)} \sum_{j\in\partial_i}\delta_{k^\prime,k_j(\bc)}}{
N\sum_n p(n|\bc)k(n)}
= \frac{\sum_{i}\sum_n \delta_{n,n_i(\bc)}\delta_{k,k(n)} \sum_{s\leq k(n)} \delta_{k^\prime,\xi^s}}{
N\sum_n p(n|\bc)k(n)}
\nonumber
\\
&=& 
\frac{\sum_n p(n|\bc)\delta_{k,k(n)} \sum_{s\leq k(n)} \delta_{k^\prime,\xi^s}}{
\sum_n p(n|\bc)k(n)}
\end{eqnarray}
Given the symmetry of $W(k,k^\prime|\bc)$ under permutation of $k$ and $k^\prime$ we then also have
\begin{eqnarray}
W(k,k^\prime|\bc)&=&
\frac{\sum_n p(n|\bc)\delta_{k^\prime,k(n)} \sum_{s\leq k(n)} \delta_{k,\xi^s}}{
\sum_n p(n|\bc)k(n)}
\label{eq:Winn}
\end{eqnarray}
The converse of the above statements is not true. One cannot calculate the neighbourhoud distribution $p(n|\bc)$ from  $p(k,m|\bc)$ or from $W(k,k^\prime|\bc)$ (or both). Note that by definition (and since $\bc$ is nondirected) we always have $W(k,k^\prime|\bc)=W(k^\prime,k|\bc)$. 

\subsection{Decomposition of graphs into directed  degree-regular subgraphs}

Any nondirected graph $\bc$ can always be decomposed uniquely into a collection of non-overlapping $N$-node subgraphs $\bbeta^{kk^\prime}$, with $k,k^\prime\in\N$, which share the nodes $\{1,\ldots,N\}$ of $\bc$ but not all of the links. These subgraphs are defined for each $(k,k^\prime)$ by the adjacency matrices
\begin{eqnarray}
\beta^{kk^\prime}_{ij}&=& c_{ij}\delta_{k,k_i(\bc)}\delta_{k^\prime,k_j(\bc)}
\label{eq:betas}
\end{eqnarray}
Each graph $\bbeta^{kk^\prime}$ contains those links in $\bc$ that go from a node with degree $k^\prime$ to a node with degree $k$. Clearly, all graphs $\bbeta^{kk^\prime}$ follow uniquely from $\bc$ via (\ref{eq:betas}). 
The converse uniqueness of $\bc$, given the matrices $\bbeta^{kk^\prime}$, is a consequence of the simple identity
\begin{eqnarray}
c_{ij}&=& c_{ij}\sum_{kk^\prime\geq 0}\delta_{k,k_i(\bc)}\delta_{k^\prime,k_j(\bc)}
=\sum_{kk^\prime\geq 0}\delta_{k,k_i(\bc)}\delta_{k^\prime,k_j(\bc)}c_{ij}
=\sum_{kk^\prime\geq 0}\beta^{kk^\prime}_{ij}
\end{eqnarray}
The graph $\bbeta^{kk^\prime}$ is directed if $k\!\neq\! k^\prime$, and nondirected  if $k\!=\! k^\prime$.
From the symmetry of $\bc$ it follows moreover that $\beta^{kk^\prime}_{ji}=\beta^{k^\prime k}_{ij}$ for all $(i,j,k,k^\prime)$, so $\bbeta^{k^\prime k}$ is specified in full by $\bbeta^{kk^\prime}$.
Although each $\bbeta^{kk^\prime}$ is an $N$-node graph, most of the nodes in $\bbeta^{kk^\prime}$ will be isolated:
 all nodes whose degrees in the original graph $\bc$ were neither $k$ nor $k^\prime$ will have degree zero in $\bbeta^{kk^\prime}$.

We now  inspect the degree statistics of the decomposition graphs $\bbeta^{kk^\prime}$, and their relation with the structural features of $\bc$. 
If $k\neq k^\prime$ we find for the remaining degrees in  $\bbeta^{kk^\prime}$:
\begin{eqnarray}
k_i(\bc)=k:&~~~~~& k^{\rm in}_i(\bbeta^{kk^\prime})=\sum_{j\in\partial_i}\delta_{k^\prime,k_j(\bc)},~~~~
k^{\rm out}_i(\bbeta^{kk^\prime})=0
\\
k_j(\bc)=k^\prime:&~~~~~& k_j^{\rm out}(\bbeta^{kk^\prime})=\sum_{i\in\partial_j}\delta_{k,k_i(\bc)},
~~~~k_j^{\rm in}(\bbeta^{kk^\prime})=0
\end{eqnarray}
Hence the joint in-out degree distribution of $\bbeta^{kk^\prime}$ can be writen in terms of the empirical distribution of neighbourhoods of $\bc$, viz. $p(n|\bc)=N^{-1}\sum_i \delta_{n,n_i(\bc)}$ with $n=(k;~\{\xi^s\})$:
\begin{eqnarray}
p^{kk^\prime}\!(q^{\rm in}\!,q^{\rm out})&=&\frac{1}{N}\sum_i \delta_{q^{\rm in},k^{\rm in}_i(\bbeta^{kk^\prime})}
\delta_{q^{\rm out},k^{\rm out}_i(\bbeta^{kk^\prime})}
\nonumber
\\
&=& \frac{1}{N}\sum_i \delta_{q^{\rm in},\delta_{k,k_i(\bc)}\sum_{j\in\partial_i}\delta_{k^\prime,k_j(\bc)}}
\delta_{q^{\rm out},\delta_{k^\prime,k_i(\bc)}\sum_{j\in\partial_i}\delta_{k,k_j(\bc)}}
\nonumber
\\
&=& \frac{1}{N}\sum_i \Big[
\delta_{k,k_i(\bc)}
\delta_{q^{\rm in},\sum_{j\in\partial_i}\delta_{k^\prime,k_j(\bc)}}
+(1\!-\!\delta_{k,k_i(\bc)})
\delta_{q^{\rm in},0}
\Big]
\nonumber
\\
&&\hspace*{10mm}
\times
\Big[
\delta_{k^\prime,k_i(\bc)}
\delta_{q^{\rm out},\sum_{j\in\partial_i}\delta_{k,k_j(\bc)}}
+
(1\!-\!\delta_{k^\prime,k_i(\bc)})
\delta_{q^{\rm out},0}
\Big]
\nonumber
\\
&=&  \sum_n p(n|\bc)\Big[
\delta_{k,k(n)}
\delta_{q^{\rm in},\sum_{s\leq k(n)}\delta_{k^\prime,\xi^s(n)}}
+(1\!-\!\delta_{k,k(n)})
\delta_{q^{\rm in},0}
\Big]
\nonumber
\\
&&\hspace*{10mm}
\times
\Big[
\delta_{k^\prime,k(n)}
\delta_{q^{\rm out},\sum_{s\leq k(n)}\delta_{k,\xi^s(n)}}
+
(1\!-\!\delta_{k^\prime,k(n)})
\delta_{q^{\rm out},0}
\Big]
\label{eq:ddistA}
\end{eqnarray}
The two marginals of (\ref{eq:ddistA}) are
\begin{eqnarray}
p_{\rm in}^{kk^\prime}(q)
&=&
 \sum_n p(n|\bc)\Big[
\delta_{k,k(n)}
\delta_{q,\sum_{s\leq k(n)}\delta_{k^\prime,\xi^s(n)}}
+(1\!-\!\delta_{k,k(n)})
\delta_{q,0}
\Big]
\\
p_{\rm out}^{kk^\prime}(q)
&=&
 \sum_n p(n|\bc)
\Big[
\delta_{k^\prime,k(n)}
\delta_{q,\sum_{s\leq k(n)}\delta_{k,\xi^s(n)}}
+
(1\!-\!\delta_{k^\prime,k(n)})
\delta_{q,0}
\Big]
\end{eqnarray}
Hence $p_{\rm in}^{kk^\prime}(q)=p_{\rm out}^{k^\prime k}(q)$, as expected.
The average degree $\bar{q}^{kk^\prime}=\sum_{q^{\rm in},q^{\rm out}}q^{\rm in}p^{kk^\prime}\!(q^{\rm in}\!,q^{\rm out})
=\sum_{q^{\rm in},q^{\rm out}}q^{\rm out}p^{kk^\prime}\!(q^{\rm in}\!,q^{\rm out})$ of the graph $\bbeta^{kk^\prime}$ can be written, using identity (\ref{eq:Winn}) and the symmetry of $W(k,k^\prime|\bc)$, as
\begin{eqnarray}
\bar{q}^{kk^\prime}&=&
\sum_n p(n|\bc)
\delta_{k(n),k}\sum_{s\leq k(n)}\delta_{k^\prime,\xi^s(n)}~=~\bar{k}(\bc)W(k,k^\prime|\bc)
\end{eqnarray}
If $k=k^\prime$, the decomposition matrix $\bbeta^{kk^\prime}$ is symmetric. Here we find
\begin{eqnarray}
k_i(\bbeta^{kk})&=&\delta_{k,k_i(\bc)}\sum_{j\in\partial_i}\delta_{k,k_j(\bc)}
\end{eqnarray}
Hence the degree distribution of $\bbeta^{kk}$ becomes
\begin{eqnarray}
p^{kk}(q)&=&
\frac{1}{N}\sum_i \delta_{q,\delta_{k,k_i(\bc)}\sum_{j\in\partial_i}\delta_{k,k_j(\bc)}}
\nonumber
\\
&=& \frac{1}{N}\sum_i \Big[
\delta_{k,k_i(\bc)}
\delta_{q,\sum_{j\in\partial_i}\delta_{k,k_j(\bc)}}
+
(1\!-\!\delta_{k,k_i(\bc)})
\delta_{q,0}\Big]
\nonumber
\\
&=&  \sum_{n} p(n|\bc)\Big[
\delta_{k,k(n)}
\delta_{q,\sum_{s\leq k(n)}\delta_{k,\xi^s(n)}}
+
(1\!-\!\delta_{k,k(n)})
\delta_{q,0}\Big]
\label{eq:ddistB}
\end{eqnarray}
The average degree in $\bbeta^{kk}$ is therefore 
\begin{eqnarray}
\bar{q}^{kk}&=& 
\sum_n p(n|\bc)\delta_{k(n),k} \sum_{s\leq k(n)}\delta_{k,\xi^s(n)}~=~\bar{k}(\bc)W(k,k|\bc)
\end{eqnarray}

\section{Entropy of ensembles of bipartite graphs}

Here we calculate the leading orders in $N$ of the entropy per node (\ref{eq:S}) for ensembles of bipartite grahs with prescribed (and possibly distinct) degree distributions in the two node sets. This is not only a novel result in itself, but will also form the seed of the entropy calculation for ensembles with constrained neighboorhoods in a subsequent section. 

In a bipartite ensemble the $N$ nodes can be divided into two disjoint sets $A,B\subseteq \{1,\ldots,N\}$ such that $c_{ij}=0$ as soon as $i,j\in A$ or $i,j\in B$, leaving only links {\em between} $A$ and $B$. This constraint implies that there is a bijective mapping from the set of bipartite graphs on on $\{1,\ldots,N\}$  to the set of directed graphs on $\{1,\ldots,N\}$, defined by assigning to each bipartite link the direction of flow from $A$ to $B$. This allows us to draw upon results on directed graphs derived in \cite{Roberts11}. The directed graph  $\bc^\prime$ associated with the bipartite graph  $\bc$ would have
\begin{eqnarray}
j\in B~~{\rm or}~~i\in A:&~~~& c^\prime_{ij}=0\\
j\in A~~{\rm and}~~i\in B:&~~~& c^\prime_{ij}=c_{ij}
\end{eqnarray}
and hence the in- and out-degree sequence $\vec{\bk}=((k_1^{\rm in},k_1^{\rm out}),\ldots,k_N^{\rm in},k_N^{\rm out}))$ of $\bc^\prime$ can be 
expressed in terms of the degree sequence $\bk$ of $\bc$ via
\begin{eqnarray}
i\in A: &~~~& \vec{k}_i=(k_i^{\rm in},k_i^{\rm out})=(0,k_i)\\
i\in B: &~~~& \vec{k}_i= (k_i^{\rm in},k_i^{\rm out})=(k_i,0)
\end{eqnarray}
The directed graph will thus have the joint degree distribution
\begin{eqnarray}
p(q^{\rm in},q^{\rm out})&=& \frac{|A|}{N}\delta_{q^{\rm in},0}p_A(q^{\rm out})+(1\!-\!\frac{|A|}{N})p_B(q^{\rm in})\delta_{q^{\rm out},0}
\end{eqnarray}
with the degree distributions 
$p_A(k)=|A|^{-1}\sum_{i\in A}\delta_{k,k_i(\bc)}$ and  $p_B(k)=|B|^{-1}\sum_{i\in B }\delta_{k,k_i(\bc)}$ in the sets $A$ and $B$ of the bipartite graph. Our bipartite ensemble is one in which we describe the distributions $p_A(k)$ and $p_B(k)$, together with the probability $f\in[0,1]$ for a node to be in subset $A$,
and we forbid links within the sets $A$ or $B$. Conservation of links demands that the two distributions cannot be independent, but must obey  $\bar{q}=(1\!-\!f)\sum_q qp_B(q)=f\sum_q qp_A(q)$, where $\bar{q}$ is the average degree. 
Our bijective mapping to directed graphs shows that the  entropy of any bipartite ensemble can be calculated by application of (\ref{eq:S},\ref{eq:core}) to an ensemble of directed graphs, with 
 $X_i=(\tau_i,k_i)$. Here $\tau_i\in\{A,B\}$ gives the subset assigment of a node. We then find
\begin{eqnarray}
S&=&
\sum_{\btau,\bk} \Big[\prod_i p(\tau_i,k_i)\Big] S(\btau,\bk) 
-f\log f-(1\!-\!f) \log(1\!-\!f) \nonumber
\\
&&-f\sum_k p_A(k)\log p_A(k)-(1\!-\!f)\sum_k p_B(k)\log p_B(k)
\end{eqnarray}
with
\begin{eqnarray}
p(\tau,k)&=& f\delta_{\tau,A}p_A(k)+(1\!-\!f)\delta_{\tau,B} p_B(k)
\\
S(\btau,\bk)&=& 
\frac{1}{N}\log \sum_{\bc}\Big(\prod_{i, \tau_i=A} \delta_{\vec{k}_i,(0,k_i)}\Big)\Big(\prod_{i, \tau_i=B} \delta_{\vec{k}_i,(k_i,0)}\Big)
\end{eqnarray}
 The latter quantity follows from the calculation in \cite{Roberts12}, with the short-hand $\pi_{\bar{q}}(q)=\rme^{-\bar{q}}\bar{q}^q/q!$ and modulo terms that vanish for $N\to\infty$:
\begin{eqnarray}
\hspace*{-5mm}
S(\btau,\bk)&=& 
\bar{q}[\log(N/\bar{q})\!+\!1]+\sum_{q}\Big[
f\delta_{q,0}+(1\!-\!f)p_B(q)\Big]\log \pi_{\bar{q}}(q)
\nonumber
\\
\hspace*{-5mm}&&
+\sum_{q}\Big[
fp_A(q)+(1\!-\!f)\delta_{q,0}\Big]\log  \pi_{\bar{q}}(q)
\nonumber
\\
&=& 
\bar{q}\log(N/\bar{q})
+f\sum_{q}
p_A(q)\log  \pi_{\bar{q}}(q)
+(1\!-\!f)\sum_{q}p_B(q)\log \pi_{\bar{q}}(q)
\hspace*{5mm}
\end{eqnarray}
This then leads to our final result for the entropy per node of tailored bipartite graph ensembles, with imposed bipartite degree distributions $p_A(k)$ and $p_B(k)$, average degree $\bar{k}$, and a fraction $f$ of nodes in the set $A$ (modulo vanishing orders in $N$): 
\begin{eqnarray}
S&=&
\bar{k}\log(N/\bar{k})
-f\log f-(1\!-\!f) \log(1\!-\!f) \nonumber
\\
&&-f\sum_k p_A(k)\log \Big(\frac{p_A(k)}{ \pi_{\bar{k}}(k)}\Big)
-(1\!-\!f)\sum_k p_B(k)\log \Big(\frac{p_B(k)}{ \pi_{\bar{k}}(k)}\Big)
\end{eqnarray}
If the sets $A$ and $B$ were to be specified explicity (as opposed to only their relative sizes), the contribution $S_f=-f\log f-(1\!-\!f) \log(1\!-\!f) $ would disappear from the above formula.

\section{Entropy of ensembles with constrained neighbourhoods}

We now turn to the Shannon entropy per node (\ref{eq:S}) of the ensemble (\ref{eq:ensemble}) in which for the observables $X_i(\bc)$ we choose the local neighbourhood $n_i(\bc)$ defined in (\ref{eq:neighbourhood}). For this we need to calculate the leading orders of
$S(\bn)=N^{-1}\log \sum_{\bc}\delta_{\bn,\bn(\bc)}$. We now use the one-to-one relationship between a graph $\bc$ and its decomposition $\bc=\sum_{qq^\prime}\bbeta^{qq^\prime}$, to write
\begin{eqnarray}
S(\bn)&=& \frac{1}{N}\log \sum_{\big\{\bbeta^{kk^\prime}\big\}}   \delta_{\bn,\bn(\bc)}
\label{eq:Sn}
\end{eqnarray}
The next argument is the key to our ability to evaluate the entropy. It involves translating the constraint $\bn=\bn(\bc)$ into constraints on the decomposition matrices $\bbeta^{kk^\prime}$. 
Let us define the sets of nodes in $\bc$ which have the same degree, viz. $I_k(\bn)=\{i\leq N|~k_i(\bc)=k\}$. 
The constraint $\bn=\bn(\bc)$ in (\ref{eq:Sn}) prescribes:
\begin{description}
\item[$~~~(i)$] all the sets $I_k$ of nodes with a given degree
\item[$~~~(ii)$] for each node $i\in I_k$ which sets $I_{k^\prime}$ this node is (possibly multiply) connected to 
\end{description}
Hence the constraint $\bn=\bn(\bc)$  specifies exactly the in- and out-degree sequences of all decomposition matrices $\bbeta^{kk^\prime}$ of $\bc$, which we will denote as $\vec{\bq}^{kk^\prime}
=(\bq^{{\rm in},kk^\prime}\!,\bq^{{\rm out},kk^\prime})$, and whose distributions we have already calculated in (\ref{eq:ddistA},\ref{eq:ddistB}). We thus see that (\ref{eq:Sn}) can be written as
\begin{eqnarray}
S(\bn)&=& \frac{1}{N}\log \sum_{\big\{\bbeta^{kk^\prime}\big\}}  \prod_{kk^\prime}\delta_{\vec{\bq}^{kk^\prime}_{\bn},\vec{\bq}(\bbeta^{kk^\prime})}
\end{eqnarray}
in which $\vec{\bq}^{kk^\prime}_{\bn}$ are the in- and out-degree sequences that are imposed by the local environment sequence $\bn$ on the decomposition matrix $\bbeta^{kk^\prime}$, and whose distributions are known to be  (\ref{eq:ddistA},\ref{eq:ddistB}). Using the symmetry $(\bbeta^{kk^\prime})^\dag=\bbeta^{k^\prime k}$ we may now write
\begin{eqnarray}
S(\bn)&=& \frac{1}{N}\log \Big[\Big(\prod_{k<k^\prime}\sum_{\bbeta^{kk^\prime}} \delta_{\vec{\bq}^{kk^\prime}_{\bn},\vec{\bq}(\bbeta^{kk^\prime})}\Big)
\Big(\prod_{k}\sum_{\bbeta^{kk}} \delta_{\vec{\bq}^{kk}_{\bn},\vec{\bq}(\bbeta^{kk})}\Big)\Big]
\nonumber
\\
&=& 
\sum_{k<k^\prime}\Big\{\frac{1}{N}\log\sum_{\bbeta^{kk^\prime}} \delta_{\vec{\bq}^{kk^\prime}_{\bn},\vec{\bq}(\bbeta^{kk^\prime})}\Big\}
+\sum_k \Big\{
\frac{1}{N}\log \sum_{\bbeta^{kk}} \delta_{\bq^{kk}_{\bn},\bq(\bbeta^{kk})}\Big\}
\label{eq:decomposedS}
\end{eqnarray}
We see that the entropy $S(\bn)$ can be written as the sum of the entropies of sub-ensembles, which are the decomposition matrices $\bbeta^{kk^\prime}$ with prescribed degree sequences. 
The second sum in (\ref{eq:decomposedS}) is over nondirected ensembles, the first over directed ones. 
The sub-entropies  were all calculated, respectively, in \cite{Annibale09} and \cite{Roberts11}\footnote{In \cite{Annibale09,Roberts11} the entropies  were carried out for ensembles with prescribed degree distributions, but it was shown that, in analogy with (\ref{eq:S}), this is simply the sum of the Shannon entropy of the degree distributions and the entropy of the corresponding ensemble with prescribed sequences.}.
The  entropy of an $N$-node nondirected random graph ensemble with degree sequence $\bq$ was found to be
 (modulo terms that vanish for $N\to\infty$):
\begin{eqnarray}
S_{\bq}&=&\frac{1}{N}\log\sum_{\bc}\delta_{\bq,\bq(\bc)}
~=~ \frac{1}{2}\bar{q}[\log (N/\bar{q})+1]+\sum_{q}p(q)\log\pi_{\bar{q}}(q)
\label{eq:entrop_nondirected}
\end{eqnarray}
in which $\bar{q}=N^{-1}\sum_i q_i$ and $\pi_{\bar{q}}(q)$ is the Poisson distribution with average $\bar{q}$.
The  entropy of an $N$-node directed random graph ensemble with in- and out-degree sequence $\vec{\bq}$ was found to be
 (modulo terms that vanish for $N\to\infty$):
\begin{eqnarray}
S_{\vec{\bq}}&=&\frac{1}{N}\log\sum_{\bc}\delta_{\vec{\bq},\vec{\bq}(\bc)} \nonumber
\\
&=& \bar{q}[\log (N/\bar{q})+1]+\sum_{q^{\rm in},q^{\rm out}}p(q^{\rm in}\!,q^{\rm out})\log[\pi_{\bar{q}}(q^{\rm in}) \pi_{\bar{q}}(q^{\rm out})]
\label{eq:entrop_directed}
\end{eqnarray}
The above entropies depend in leading orders only on the degree distributions (as opposed to the degree sequences), and since these distributions were  already calculated (\ref{eq:ddistA},\ref{eq:ddistB}), 
we can simply insert (\ref{eq:entrop_nondirected},\ref{eq:entrop_directed}) into (\ref{eq:decomposedS}), with the correct distributions (\ref{eq:ddistA},\ref{eq:ddistB}), and find an expression that depends only on the local environment distribution $p(n)=N^{-1}\sum_i \delta_{n,n_i}$:
\begin{eqnarray}
\hspace*{-5mm}
S(\bn)&=& 
\sum_{k<k^\prime}\Big\{
 \bar{q}^{kk^\prime}[\log (N/\bar{q}^{kk^\prime})\!+\!1]+\sum_{q^{\rm in},q^{\rm out}}p^{kk^\prime}\!(q^{\rm in}\!,q^{\rm out})\log[\pi_{\bar{q}^{kk^\prime}}(q^{\rm in}) \pi_{\bar{q}^{kk^\prime}}(q^{\rm out})]
\Big\}
\nonumber
\\&&
+\sum_k \Big\{
\frac{1}{2}\bar{q}^{kk}[\log (N/\bar{q}^{kk})\!+\!1]+\sum_{q}p^{kk}(q)\log\pi_{\bar{q}^{kk}}(q)
\Big\}
\nonumber
\\
&=& \frac{1}{2}
\sum_{k\neq k^\prime}\Big\{
 \bar{k}W(k,k^\prime)[\log (N/\bar{k}W(k,k^\prime))\!-\!1]
\nonumber
\\
&&
+2\bar{k}W(k,k^\prime)\log[\bar{k}W(k,k^\prime)]
-\sum_{q}[p^{kk^\prime}_{\rm in}(q)+p^{kk^\prime}_{\rm out}(q)]\log q!
\Big\}
\nonumber
\\&&
+\frac{1}{2}\sum_k \Big\{
\bar{k}W(k,k)[\log (N/\bar{k}W(k,k))\!-\!1]
\nonumber
\\
&&
+2\bar{k}W(k,k)\log[\bar{k}W(k,k^\prime)]
-2\sum_{q}p^{kk}(q)\log q!
\Big\}
\nonumber
\\
&=& \frac{1}{2}
 \bar{k}[\log (N/\bar{k})-1]
+\frac{1}{2}\bar{k}\sum_{k,k^\prime}W(k,k^\prime)\log W(k,k^\prime)
\nonumber
\\
&&
-\sum_q\Big[
\frac{1}{2}\sum_{k\neq k^\prime}[p^{kk^\prime}_{\rm in}(q)+p^{kk^\prime}_{\rm out}(q)]
+\sum_k p^{kk}(q)\Big]\log q!
\nonumber
\\
&=& 
\frac{1}{2}
 \bar{k}[\log (N/\bar{k})-1]
+\frac{1}{2}\bar{k}\sum_{k,k^\prime}W(k,k^\prime)\log W(k,k^\prime)
\nonumber
\\
&&
-\sum_q\sum_n p(n)
\sum_{k,k^\prime}\Big[
\delta_{k,k(n)}
\delta_{q,\sum_{s\leq k(n)}\delta_{k^\prime,\xi^s(n)}}
+(1\!-\!\delta_{k^\prime,k(n)})
\delta_{q,0}
\Big]\log q!
\nonumber
\\
&=& 
\frac{1}{2}
 \bar{k}[\log (N/\bar{k})-1]
+\frac{1}{2}\bar{k}\sum_{k,k^\prime}W(k,k^\prime)\log W(k,k^\prime)
\nonumber
\\
&&
-\sum_n p(n)
\sum_{k}
\log \Big[\Big(\sum_{s\leq k(n)}\delta_{k,\xi^s(n)}\Big)!\Big] 
\end{eqnarray}
Insertion of this result into the general formula (\ref{eq:S}) gives us an analytical expression for the Shannon entropy of the random graph ensemble with prescribed distribution $p(n)$ of local neighbourhoods, modulo terms that vanish for  $N\to\infty$. This expression is fully explicit, since  $\bar{k}$ and $W(k,k^\prime)$ are both determined by the distribution $p(n)$, via $\bar{k}=\sum_n p(n)k(n)$ and 
(\ref{eq:Winn}) respectively:
\begin{eqnarray}
S&=& 
\frac{1}{2}
 \bar{k}[\log (N/\bar{k})-1]
+\frac{1}{2}\bar{k}\sum_{k,k^\prime}W(k,k^\prime)\log W(k,k^\prime)
\nonumber
\\
&&
-\sum_{n}p(n)\log p(n)
-\sum_n p(n)
\sum_{k}
\log \Big[\Big(\sum_{s\leq k(n)}\delta_{k,\xi^s(n)}\Big)!\Big] 
\label{eq:final_entropy_neighbourhoods}
\end{eqnarray}

\section{Entropy of ensembles of networks with specified generalized degree distribution}
\label{sec:generalized_degrees_direct_calculation}

In this section we consider an ensemble of nondirected networks with a specified generalized degree distribution $p(k,m)=N^{-1}\sum_i \delta_{k,k_i(\bc)}\delta_{m,m_i(\bc)}$, where  $k_i(\bc) = \sum_j c_{ij}$ and $m_i = \sum_{jk}c_{ij} c_{jk}$. Previous work \cite{bianconi2008entropies} began this calculation, and reached (in leading order) the intermediate form set out below:
\begin{eqnarray}
S &= & \frac{1}{2}\bar{k}[ \log(N/\bar{k}) \!+\!1]-\sum_{k,m} p(k,m) \log\Big(\frac{p(k,m)}{\pi(k)}  \Big)
\nonumber
\\ 
&&\hspace*{20mm}
+\sum_{k,m}p(k,m) \log \Big(\sum_{\xi^1,.,\xi^k}  \delta_{m, \sum^k_{s=1} \xi^s}\prod_{s=1}^k \gamma(k, \xi^s) \Big)
\label{eq:intermediate_form}
\end{eqnarray}
  $\bar{k}$ indicates the average degree; $\pi_{\bar{k}}(k)$ is the Poissonian distribution with  average degree $\bar{k}$. The sum inside the logarithm in the final term of (\ref{eq:intermediate_form}) runs over all sets of $k$ nonnegative integers $\xi^1\ldots \xi^k$. The function $\gamma(.,.)$ is defined as the non-negative solution to the following self-consistency relation:
\begin{eqnarray}
\label{eq:gd_gamma}
\gamma({k, k^\prime}) 
= 
\sum_{m^\prime} \frac{k^\prime}{\bar{k}} \, p({k}^\prime, {m}^\prime) 
\left[
	\frac
	    {
	    	\sum_{\xi^1...\xi^{k^\prime-1}} 
	    		\delta_{m^\prime-k, \sum_{s=1}^{k^\prime-1}\xi^s}
	    		\prod_{s=1}^{k^\prime-1} \gamma\left( k^\prime, \xi^s \right)
	    }
	    {
	    	\sum_{\xi^1...\xi^{k^\prime}} 
	    		\delta_{m^\prime, \sum_{s=1}^{k^\prime}\xi^s}
	    		\prod_{s=1}^{k^\prime} \gamma\left( k^\prime, \xi^s \right)
	    }
\right]
\end{eqnarray}
This equation does not yield to a straightforward solution, and can only be evaluated numerically or in certain special cases. Without a physical interpretation of $\gamma(k,k^\prime)$, this intermediate answer is limited in how much insight it can provide. We will now show how the entropy can be expressed in terms of measurable quantities.

Our strategy is to derive an expression for the (observable) degree-degree correlations $W(k,k^\prime)$, and show that these can be expressed it terms of the order parameter $\gamma(k,k^\prime)$ that appears in equation \eref{eq:intermediate_form}. 
We calculate the average of this quantity in our tailored ensembles of the form (\ref{eq:ensemble}), where we now define topological characteristics by specifying a generalised degree distribution $p(k,m)$.
We follow closely the steps taken in \cite{bianconi2008entropies}, and write for the ensemble a specified generalised degree sequence $(\bk,\bm)$:
\begin{eqnarray}
\hspace*{-25mm}
W(k,k^\prime)&=&\frac{1}{N\overline{k}}\sum_{\bc}p(\bc|\bk,\bm)\sum_{rs} c_{rs}\delta_{k,\sum_\ell c_{r\ell}}\delta_{k^\prime,\sum_\ell c_{s\ell}}
\nonumber
\\
\hspace*{-25mm}
&&\hspace*{-10mm}=\frac{1}{N^2}\sum_{rs}\delta_{k,k_r}\delta_{k^\prime\!,k_s}
 \frac{\int_{\pi}^\pi\!\rmd\btheta \rmd\bphi~\rme^{\rmi(\btheta\cdot\bk+\bphi\cdot\bm)-\rmi(\theta_r+\theta_s+\phi_r k_s+\phi_s k_r)}
\prod_{i<j} \Big[1\!+\!
\frac{\overline{k}}{N}\Big(\rme^{-\rmi(\theta_i+\theta_j+\phi_ik_j+\phi_j k_i)}\!-\!1\Big)
\Big]
}
{ \int_{\pi}^\pi\!\rmd\btheta \rmd\bphi~\rme^{\rmi(\btheta\cdot\bk+\bphi\cdot\bm)}
\prod_{i<j} \Big[1\!+\!
\frac{\overline{k}}{N}\Big(e^{-i(\theta_i+\theta_j+\phi_ik_j+\phi_j k_i)}\!-\!1\Big)
\Big]}
\nonumber
\\[-1mm]
\hspace*{-25mm}&&\hspace*{95mm}
+\order(\frac{1}{N})
\nonumber
\\
\hspace*{-25mm}
&&\hspace*{-10mm}
=\frac{1}{N^2}\sum_{rs}\delta_{k,k_r}\delta_{k^\prime\!,k_s} \frac{\int_{\pi}^\pi\!\rmd\btheta \rmd\bphi~\rme^{\rmi(\btheta\cdot\bk+\bphi\cdot\bm)-\rmi(\theta_r+\theta_s+\phi_r k^\prime+\phi_s k)+\frac{\overline{k}}{2N}\sum_{ij}\rme^{-\rmi(\theta_i+\theta_j+\phi_ik_j+\phi_j k_i)}+\ldots}
}
{ \int_{\pi}^\pi\!\rmd\btheta \rmd\bphi~\rme^{\rmi(\btheta\cdot\bk+\bphi\cdot\bm)
+\frac{\overline{k}}{2N}\sum_{ij}
\rme^{-\rmi(\theta_i+\theta_j+\phi_ik_j+\phi_j k_i)}+\ldots}}
+\order(\frac{1}{N})
\nonumber
\\
\hspace*{-25mm}
&&\hspace*{-10mm}
=\frac{\int_{\pi}^\pi\!\rmd\btheta \rmd\bphi~\rme^{\rmi(\btheta\cdot\bk+\bphi\cdot\bm)+\frac{\overline{k}}{2N}\sum_{ij}\rme^{-\rmi(\theta_i+\theta_j+\phi_ik_j+\phi_j k_i)}}
\Big(\frac{1}{N^2}\sum_{rs}\delta_{k,k_r}\delta_{k^\prime\!,k_s} \rme^{-\rmi(\theta_r+\theta_s+\phi_r k^\prime+\phi_s k)}\Big)
}
{ \int_{\pi}^\pi\!\rmd\btheta \rmd\bphi~\rme^{\rmi(\btheta\cdot\bk+\bphi\cdot\bm)
+\frac{\overline{k}}{2N}\sum_{ij}
\rme^{-\rmi(\theta_i+\theta_j+\phi_ik_j+\phi_j k_i)}}}
+\order(\frac{1}{N})\hspace*{-5mm}
\nonumber
\\
\hspace*{-25mm}
&&\hspace*{-15mm}
=\frac{\int_{\pi}^\pi\!\rmd\btheta \rmd\bphi~\rme^{\rmi(\btheta\cdot\bk+\bphi\cdot\bm)+\frac{\overline{k}}{2N}\sum_{ij}\rme^{-\rmi(\theta_i+\theta_j+\phi_ik_j+\phi_j k_i)}}
\Big(
\frac{1}{N}\sum_{r}\delta_{k,k_r}\rme^{-\rmi(\theta_r+\phi_r k^\prime)}\Big)
\Big(
\frac{1}{N}\sum_{s}\delta_{k^\prime\!,k_s} \rme^{-\rmi(\theta_s+\phi_s k)}\Big)
}
{ \int_{\pi}^\pi\!\rmd\btheta \rmd\bphi~\rme^{\rmi(\btheta\cdot\bk+\bphi\cdot\bm)
+\frac{\overline{k}}{2N}\sum_{ij}
\rme^{-\rmi(\theta_i+\theta_j+\phi_ik_j+\phi_j k_i)}}}
+\order(\frac{1}{N})
\nonumber
\\
\hspace*{-25mm}&&\hspace*{-10mm}=\frac{\int\{\rmd P \rmd\hat{P}\} \rme^{N\Psi[P,\hat{P}]}
\Big(\int\!\rmd\theta \rmd\phi~P(\theta,\phi,k)\rme^{-\rmi\theta-\rmi\phi k^\prime}\Big)
\Big(\int\!\rmd\theta \rmd\phi~P(\theta,\phi,k^\prime)\rme^{-\rmi\theta-\rmi\phi k}\Big)
}
{ \int\{\rmd P \rmd\hat{P}\} \rme^{N\Psi[P,\hat{P}]}
}
+\order(\frac{1}{N})
\end{eqnarray}
Taking the limit $N\to\infty$ therefore gives
\begin{eqnarray}
\hspace*{-20mm}
\lim_{N\to\infty}W(k,k^\prime)
&=&
 \Big(\int\!\rmd\theta \rmd\phi~P(\theta,\phi,k)\rme^{-\rmi\theta-\rmi\phi k^\prime}\Big)
\Big(\int\!\rmd\theta \rmd\phi~P(\theta,\phi,k^\prime)\rme^{-\rmi\theta-\rmi\phi k}\Big)\Big|_{{\rm saddle-point}~\{P,\hat{P}\}~{\rm of}~\Psi}
\end{eqnarray}
in which the function $\Psi[P,\hat{P}]$ is identical to that found in \cite{bianconi2008entropies}. Using the 
the formulae in \cite{bianconi2008entropies} that relate to the definition of the order parameter $\gamma(k,k^\prime)$, we then 
obtain for $N\to\infty$  the unexpected simple but welcome relation 
\begin{eqnarray}
W(k,k^\prime)&=& \gamma(k,k^\prime)\gamma(k^\prime,k)
\label{eq:W_and_gamma}
\end{eqnarray}
A similar, although slightly more involved, calculation leads to an expression for the joint distribution $W(k,m;k^\prime,m^\prime)$; see the Appendix for details.

Our final  aim is to use identity (\ref{eq:W_and_gamma}) to resolve equation \eref{eq:intermediate_form} into observable quantities. Consider the nontrivial term in (\ref{eq:intermediate_form}):
\begin{eqnarray}
 \Gamma &=& 
\sum_{k,m}p(k,m) \log \Big(\sum_{\xi^1,.,\xi^k} \!\prod_{s=1}^k \gamma(k, \xi^s) \delta_{m, \sum^k_{s=1} \xi^s} \Big)
 \end{eqnarray}
At this point  of the calculation, the effect of factorising across nodes has been to break the expression down into terms which, for every generalised degree $(k,m)$, enumerate all the possible ways of dividing $m$ second neighbours between $k$ first neighbours. The term inside the logarithm sums for each $k$ over all configurations $\{\xi^1\ldots \xi^k\}$ which meet the condition $\sum_{s=1}^k \xi^s = m$. To formalise this idea, we may re-aggregate the expression for any graphically realisable distribution $p(k,m)$ to write
\begin{eqnarray}
 \Gamma &=& \frac{1}{N}
\log \Big\{\prod_{k,m}\Big(\sum_{\xi^1,.,\xi^k} \!\prod_{s=1}^k \gamma(k, \xi^s) \delta_{m, \sum^k_{s=1} \xi^s} \Big)^{Np(k,m)}
\nonumber
\\
&=& 
\frac{1}{N}
\log \prod_i \Big(\sum_{\xi_i^1\ldots \xi_i^{k_i}} \!\Big[\prod_{s=1}^{k_i} \gamma(k_i, \xi_i^{s})\Big] \delta_{m_i, \sum^{k_i}_{s=1} \xi_i^s} \Big)
\nonumber
\\
&=& 
\frac{1}{N}
\log\Big\{\sum_{\xi_1^1\ldots \xi_1^{k_1}}\ldots \sum_{\xi_N^1\ldots \xi_N^{k_N}}\Big(\prod_i \delta_{m_i, \sum^{k_i}_{s=1} \xi_i^s}\Big)
 \prod_i \prod_{s=1}^{k_i} \gamma(k_i, \xi_i^{s})\Big\}
\label{eq:bigsum}
 \end{eqnarray}
We can now see that the separate terms precisely enumerate all the permutations of degrees and neighbour-degrees for networks with a generalised degree sequence consistent with any pair $(k,m)$ appearing $N p(k,m)$ times. The Kronecker deltas $\delta_{m_i, \sum^{k_i}_{s=1} \xi_i^s} $ tell us that each $\xi_i^s$ 
in any nonzero term is to be interpreted as the degree of a node $j\in \partial_i$, and must therefore appear also as the left argument in another factor of the type $\gamma(k_j,.)$. 
This insight allows the expression to be substantially simplified, since we already know that $\gamma(k, k^\prime)\gamma(k^\prime, k)= W(k^\prime, k)$ where $W(k^\prime, k)$ is the correlation between degrees of connected nodes. Hence, any nonvanishing contribution to the sum over all neighbourhoods inside the logarithm of (\ref{eq:bigsum}) will be equal to a repeated product of factors $W(k,k^\prime)$, with different $(k,k^\prime)$. 
Since we also know that the number of links between nodes with degree combination $(k,k^\prime)$ equals $N\bar{k}W(k,k^\prime)$ in leading order in $N$, 
we conjecture that in leading order we may make the following replacement inside (\ref{eq:bigsum}):
\begin{eqnarray}
 \prod_i \prod_{s=1}^{k_i} \gamma(k_i, \xi_i^{s}) &~~\to~~& \prod_{k, k^\prime} W(k, k^\prime)^{\frac{\bar{k}N}{2} W(k, k^\prime)}
\end{eqnarray}
(where the factor $\frac{1}{2}$ in the exponent reflects the fact that two $\gamma(.,.)$ factors combine to form each factor  $W(.,.)$). 
With this conjecture we obtain, in leading order in $N$:
\begin{eqnarray}
\Gamma&=&
\frac{1}{N}
\log\Big\{\sum_{\xi_1^1\ldots \xi_1^{k_1}}\ldots \sum_{\xi_N^1\ldots \xi_N^{k_N}}\Big(\prod_i \delta_{m_i, \sum^{k_i}_{s=1} \xi_i^s}\Big)
\Big\}
\nonumber
\\
&& 
+\frac{1}{2}\bar{k} 
\sum_{k, k^\prime}W(k, k^\prime)
\log   W(k, k^\prime)
\nonumber
\\
&=&
\sum_{k,m}p(k,m) \log \Big(\sum_{\xi^1,.,\xi^k} \delta_{m, \sum^k_{s=1} \xi^s}  \Big)
+\frac{1}{2}\bar{k} 
\sum_{k, k^\prime}W(k, k^\prime)
\log   W(k, k^\prime)
 \end{eqnarray}
This implies that (\ref{eq:intermediate_form}) simplifies to
\begin{eqnarray}
S &= & \frac{1}{2}\bar{k}[ \log(N/\bar{k}) \!+\!1]
+\frac{1}{2}\bar{k} 
\sum_{k, k^\prime}W(k, k^\prime)
\log   W(k, k^\prime)
\nonumber
\\ 
&&
-\sum_{k,m} p(k,m) \log\Big(\frac{p(k,m)}{\pi(k)}  \Big)
+\sum_{k,m}p(k,m) \log \Big(\sum_{\xi^1,.,\xi^k} \delta_{m, \sum^k_{s=1} \xi^s}  \Big)
\label{eq:final_answer}
\end{eqnarray}

\section{Conclusion}

Ensembles of tailored random graphs are extremely useful constructions in the modelling of complex interacting particle systems in biology, physics, computer science, economics and the social sciences. They allow us to quantify topological features of such systems and reason quantitatively about their complexity, as well as define and generate useful  random proxies for realistic networks in statistical mechanical analyses of processes. 

In this paper we have derived, in leading two orders in $N$, explicit expressions for the Shannon entropies of different types of tailored random graph ensembles, for which no such expressions had yet been obtained. This work builds on and extends the ideas and techniques developed in the three papers 
\cite{Annibale09,bianconi2008entropies,Roberts11}, which use path integral representations to achieve link factorisation in the various summations over graphs. We show in this paper how the new ensemble entropies can often be calculated by efficient use and combination of earlier results.

The first class of graph ensembles we studied consists of bipartite nondirected graphs with prescribed (and possibly nonidentical) distributions of degrees for the two node subsets. This case could be handled by a bijective mapping from bipartite to directed graphs, for which formulae are available. The second class consists of graphs with prescribed distributions of local neighbourhoods, where the neighbourhood of a node is defined as its own degree plus the values of the degrees of its immediate neighbours. This problem was solved using a decomposition in terms of bipartite graphs, building on the previous result. The final class of graphs, for which the entropy had in the past only partially been calculated, consist of graphs with presecribed distributions of generalised degrees, i.e. of ordinary degrees plus the total number of length-two paths starting in the specified nodes. Here we derive two novel and exact identities linking the order parameters to macroscopic observables, which lead to an explicit entropy formula based on a plausible but not yet proven conjecture, 

Since completing this work, our attention has been drawn to a preprint \cite{Caputo13} which considers the question of the entropy of random graph ensembles constrained with a given distribution of neighbourhoods by a probability theory route, via an adapted Configuration Model. In that case, the neighbourhoods were specified as graphlets of an arbitrary depth. \cite{Caputo13} also retrieves the entropy of an ensemble constrained with a specified degree distribution, as originally derived by \cite{Annibale09}. 

\section*{Acknowledgements}

ESR gratefully acknowledges financial support from the Biotechnology and Biological Sciences Research Council of the United Kingdom. 

\section*{References}

\bibliography{gd4}
\bibliographystyle{iopart-num}

\appendix

\section{Generalised degree correlation kernel for ensembles with prescribed generalised degrees}
\label{app:Whard}

The generalised quantity $W(k,m;k^\prime,m^\prime)$ in the ensemble with presecribed generalised degree distributions $p(k,m)$ can be calculated along the same lines as the calculation of $W(k,k^\prime)$ in the main text. It is defined as
\begin{eqnarray}
W(k,m;k^\prime,m^\prime|\bc)&=& \frac{1}{N\bar{k}}\sum_{ij}c_{ij}
\delta_{k,\sum_\ell c_{i\ell}}\delta_{k^\prime,\sum_\ell c_{j\ell}}
\delta_{m,\sum_\ell c_{i\ell}k_\ell}\delta_{m^\prime,\sum_\ell c_{j\ell}k_\ell}
\end{eqnarray}
and its ensemble average takes the form 
\begin{eqnarray}
\hspace*{-25mm}
W(k,m;k^\prime,m^\prime)&&
\nonumber\\
\hspace*{-25mm}&&\hspace*{-20mm}=
\frac{\int_{\pi}^\pi\!\rmd\btheta \rmd\bphi~\rme^{\rmi(\btheta\cdot\bk+\bphi\cdot\bm)+\frac{\overline{k}}{2N}\sum_{ij}\rme^{-\rmi(\theta_i+\theta_j+\phi_ik_j+\phi_j k_i)}}
\Big(\frac{1}{N^2}\sum_{rs}\delta_{k,k_r}\delta_{k^\prime\!,k_s}\delta_{m,m_r}\delta_{m^\prime,m_s} \rme^{-\rmi(\theta_r+\theta_s+\phi_r k^\prime+\phi_s k)}\Big)
}
{ \int_{\pi}^\pi\!\rmd\btheta \rmd\bphi~\rme^{\rmi(\btheta\cdot\bk+\bphi\cdot\bm)
+\frac{\overline{k}}{2N}\sum_{ij}
\rme^{-\rmi(\theta_i+\theta_j+\phi_ik_j+\phi_j k_i)}}}
+\order(\frac{1}{N})\hspace*{-3mm}
\nonumber
\\
\hspace*{-25mm}
&&\hspace*{-20mm}
=\frac{\int_{\pi}^\pi\!\rmd\btheta \rmd\bphi~\rme^{\rmi(\btheta\cdot\bk+\bphi\cdot\bm)+\frac{\overline{k}}{2N}\sum_{ij}\rme^{-\rmi(\theta_i+\theta_j+\phi_ik_j+\phi_j k_i)}}
\Big(\frac{1}{N}\sum_{r}\delta_{k,k_r}\delta_{m,m_r}\rme^{-\rmi(\theta_r+\phi_r k^\prime)}\Big)
\Big(\frac{1}{N}\sum_{s}\delta_{k^\prime\!,k_s}\delta_{m^\prime,m_s} \rme^{-\rmi(\theta_s+\phi_s k)}\Big)
}
{ \int_{\pi}^\pi\!\rmd\btheta \rmd\bphi~\rme^{\rmi(\btheta\cdot\bk+\bphi\cdot\bm)
+\frac{\overline{k}}{2N}\sum_{ij}
\rme^{-\rmi(\theta_i+\theta_j+\phi_ik_j+\phi_j k_i)}}}\hspace*{-3mm}
\nonumber
\\[-1mm]
\hspace*{-15mm}&&\hspace*{85mm}
+\order(\frac{1}{N})
\end{eqnarray}
Now we will want to introduce a generalised order parameter, namely
\begin{eqnarray}
P(\theta,\phi,k,m)=
\frac{1}{N}\sum_{r}\delta_{k,k_r}\delta_{m,m_r}\delta(\theta-\theta_r)\delta(\phi-\phi_r)
\end{eqnarray}
The previous order parameter used in he calculation of $W(k,k^\prime)$ is a marginal of this, via $P(\theta,\phi,k)=\sum_m P(\theta,\phi,k,m)$. 
This definition will give us 
\begin{eqnarray}
\hspace*{-20mm}
W(k,m;k^\prime,m^\prime)&=&\Big(\int_{-\pi}^\pi\!\rmd\theta \rmd\phi~P(\theta,\phi,k,m)\rme^{-\rmi\theta-\rmi\phi k^\prime}\Big)
\Big(\int_{-\pi}^\pi\!\rmd\theta \rmd\phi~P(\theta,\phi,k^\prime,m^\prime)\rme^{-\rmi\theta-\rmi\phi k}\Big)
\\
\hspace*{-20mm}
W(k,k^\prime)&=& \sum_{mm^\prime}W(k,m;k^\prime,m^\prime)
\end{eqnarray}
in which the new order parameter and its conjugate are to be solved by extremisation of the generalised surface
\begin{eqnarray}
\hspace*{-15mm}
\Psi[P,\hat{P}]&=& \rmi\sum_{km}\int_{-\pi}^\pi\!\rmd\theta \rmd\phi~\hat{P}(\theta,\phi,k,m)P\theta,\phi,k,m)
\nonumber
\\
\hspace*{-15mm}
&&
+\sum_{km}P(k,m)\log\int_{-\pi}^\pi \!\rmd\theta \rmd\phi~\rme^{\rmi(\theta k+\phi m-\hat{P}(\theta,\phi,k,m))}
\nonumber
\\
\hspace*{-15mm}
&&+\frac{1}{2}\bar{k}\int\!\rmd\theta \rmd\phi \rmd\theta^\prime \rmd\phi^\prime \sum_{kk^\prime m m^\prime}
P(\theta,\phi,k,m)P(\theta^\prime,\phi^\prime,k^\prime,m^\prime)\rme^{-\rmi(\theta+\theta^\prime+\phi k^\prime+\phi^\prime k)}
\end{eqnarray}
Variation of $\Psi$ gives the following saddle-point equations
\begin{eqnarray}
\hat{P}(\theta,\phi,k,m)&=&
\rmi\bar{k}\rme^{-\rmi\theta}\int\! \rmd\theta^\prime \rmd\phi^\prime \sum_{k^\prime m^\prime}
P(\theta^\prime,\phi^\prime,k^\prime,m^\prime)\rme^{-\rmi(\theta^\prime+\phi k^\prime+\phi^\prime k)}
\\
 P(\theta,\phi,k,m)
&=&
P(k,m)
\frac{\rme^{\rmi(\theta k+\phi m-\hat{P}(\theta,\phi,k,m))}}
{\int_{-\pi}^\pi \!\rmd\theta^\prime \rmd\phi^\prime ~ \rme^{\rmi(\theta^\prime k+\phi^\prime m-\hat{P}(\theta^\prime\!,\phi^\prime,k,m))}}
\end{eqnarray}
Clearly $\hat{P}(\theta,\phi,k,m)=\hat{P}(\theta,\phi,k)$ (i.e. it is independent of $m$). 
We may therefore substitute $\hat{P}(\theta,\phi,k)=\rmi\bar{k}\rme^{-\rmi\theta}\hat{P}(\phi,k)$ and find
\begin{eqnarray}
\hat{P}(\phi,k)&=&
\int\! \rmd\theta^\prime \rmd\phi^\prime \sum_{k^\prime m^\prime}
P(\theta^\prime,\phi^\prime,k^\prime,m^\prime)\rme^{-\rmi(\theta^\prime+\phi k^\prime+\phi^\prime k)}
\\
 P(\theta,\phi,k,m)
&=&
P(k,m)
\frac{\rme^{\rmi(\theta k+\phi m)+\bar{k}\rme^{-\rmi\theta}\hat{P}(\phi,k))}}
{\int_{-\pi}^\pi \!\rmd\theta^\prime \rmd\phi^\prime ~ \rme^{\rmi(\theta^\prime k+\phi^\prime m)+\bar{k}\rme^{-\rmi\theta^\prime}\hat{P}(\phi^\prime,k))}}
\end{eqnarray}
We observe as before in \cite{bianconi2008entropies} that 
\begin{eqnarray}
\int_{-\pi}^\pi \!\rmd\theta~P(\theta,\phi,k)\rme^{-\rmi\theta}&=& \sum_m
P(k,m)
\frac{\int_{-\pi}^\pi \!\rmd\theta~\rme^{\rmi(\theta (k-1)+\phi m)+\bar{k}\rme^{-\rmi\theta}\hat{P}(\phi,k))}}
{\int_{-\pi}^\pi \!\rmd\theta \rmd\phi^\prime ~ \rme^{\rmi(\theta k+\phi^\prime m)+\bar{k}\rme^{-\rmi\theta}\hat{P}(\phi^\prime,k))}}
\nonumber
\\
&=& \sum_m
P(k,m)
\frac{\sum_{\ell\geq 0}\frac{\bar{k}^\ell\hat{P}^\ell(\phi,k)}{\ell!}
\int_{-\pi}^\pi \!\rmd\theta~\rme^{\rmi(\theta (k-1-\ell)+\phi m)}}
{\sum_{\ell\geq 0}\frac{\bar{k}^\ell \hat{P}^\ell(\phi^\prime,k)}{\ell!}
\int_{-\pi}^\pi \!\rmd\theta \rmd\phi^\prime ~ \rme^{\rmi(\theta k+\phi^\prime m-\ell\theta)}}
\nonumber
\\
&=& \sum_m
P(k,m)
\frac{\frac{\bar{k}^{k-1}\hat{P}^{k-1}(\phi,k)}{(k-1)!}\rme^{\rmi\phi m}}
{\frac{\bar{k}^k \hat{P}^k(\phi^\prime,k)}{k!}
\int_{-\pi}^\pi \!\rmd\phi^\prime ~ \rme^{\rmi\phi^\prime m}}
\nonumber
\\
&=& \sum_m\frac{k}{\bar{k}}
P(k,m)
\frac{\hat{P}^{k-1}(\phi,k)\rme^{\rmi\phi m}}
{
\int_{-\pi}^\pi \!\rmd\phi^\prime ~\hat{P}^k(\phi^\prime,k) \rme^{\rmi\phi^\prime m}}
\end{eqnarray}
Hence
\begin{eqnarray}
\hat{P}(\phi,k)&=&  \sum_{k^\prime m^\prime}\frac{k^\prime}{\bar{k}}
P(k^\prime,m^\prime) \rme^{-\rmi\phi k^\prime}
\frac{\int_{-\pi}^{\pi}\!  \rmd\phi^\prime~\hat{P}^{k^\prime-1}(\phi^\prime,k^\prime)\rme^{\rmi\phi^\prime( m^\prime-k)}}
{
\int_{-\pi}^\pi \!\rmd\phi^\prime ~\hat{P}^{k^\prime}(\phi^\prime,k^\prime) \rme^{\rmi\phi^\prime m^\prime}}
\end{eqnarray}
After writing $\hat{P}(\phi,k)=\sum_{k^\prime}\gamma(k,k^\prime)\rme^{-\rmi\phi k^\prime}$ we recover our familiar equation
\begin{eqnarray}
\hspace*{-10mm}
\gamma(k,k^\prime)\gamma(k^\prime,k)&=&  \frac{k^\prime}{\bar{k}}\sum_m
P(k^\prime,m) 
\frac{\sum_{k_1\ldots k_{k^\prime}} \Big[\prod_{n=1}^{k^\prime}\gamma(k^\prime,k_n)\Big]
\delta_{m,\sum_{n\leq k^\prime}k_n}\delta_{k k_n}
}
{\sum_{k_1\ldots k_{k^\prime}}
\Big[\prod_{n=1}^{k^\prime}\gamma(k^\prime,k_n)\Big]\delta_{ m,\sum_{n\leq k^\prime}k_n}}
\end{eqnarray}
But now we can also work out the generalised kernel:
\begin{eqnarray}
\hspace*{-10mm}
W(k,m;k^\prime,m^\prime)&=& \Big(\int_{-\pi}^\pi\!\rmd\theta d\phi~P(\theta,\phi,k,m)\rme^{-\rmi\theta-\rmi\phi k^\prime}\Big)
\Big(\int_{-\pi}^\pi\!\rmd\theta \rmd\phi~P(\theta,\phi,k^\prime,m^\prime)\rme^{-\rmi\theta-i\phi k}\Big)
\nonumber
\\
\hspace*{-10mm}
&&\hspace*{-20mm} =\frac{kk^\prime}{\bar{k}^2}
P(k,m)P(k^\prime,m^\prime) \Big(
\frac{\int_{-\pi}^\pi\!d\phi~\hat{P}^{k-1}(\phi,k)\rme^{\rmi\phi (m-k^\prime)}}
{
\int_{-\pi}^\pi \!\rmd\phi ~\hat{P}^k(\phi,k) \rme^{\rmi\phi m}}
\Big)
\Big(
\frac{\int_{-\pi}^\pi\!\rmd\phi~\hat{P}^{k^\prime-1}(\phi,k^\prime)\rme^{\rmi\phi (m^\prime-k)}}
{
\int_{-\pi}^\pi \!\rmd\phi ~\hat{P}^{k^\prime}(\phi,k^\prime) \rme^{\rmi\phi m^\prime}}
\Big)
\nonumber
\\
\hspace*{-10mm}
&=&\frac{kk^\prime}{\bar{k}^2}
\frac{P(k,m)P(k^\prime,m^\prime) }{\gamma(k,k^\prime)\gamma(k^\prime,k)}
\Big(
\frac{\sum_{k_1\ldots k_{k}}\Big[\prod_{n=1}^{k}\gamma(k,k_n)\Big]
\delta_{m,\sum_{n\leq k}k_n}\delta_{k_k,k^\prime}}
{\sum_{k_1\ldots k_{k}}\Big[\prod_{n=1}^{k}\gamma(k,k_n)\Big]
\delta_{m,\sum_{n\leq k}k_n}}
\Big)
\nonumber
\\
&&\times
\Big(
\frac{\sum_{k_1\ldots k_{k^\prime}}
\Big[\prod_{n=1}^{k^\prime}\gamma(k^\prime,k_n)\Big]
\delta_{m^\prime,\sum_{n\leq k^\prime}k_n}\delta_{k_{k^\prime},k}}
{\sum_{k_1\ldots k_{k^\prime}}\Big[\prod_{n=1}^{k^\prime}\gamma(k^\prime,k_n)\Big]
\delta_{ m^\prime,\sum_{n\leq k^\prime}k_n}}
\Big)
\end{eqnarray}
We  know that $W(k,k^\prime)=\gamma(k,k^\prime)\gamma(k^\prime,k)$, and that $P(k,m)k/\bar{k}=W(k,m)$,  so this can be simplified to
\begin{eqnarray}
W(k,m;k^\prime,m^\prime)&=&
\frac{W(k,m)W(k^\prime,m^\prime) }{W(k,k^\prime)}
\Big(
\frac{\sum_{k_1\ldots k_{k}}\Big[\prod_{n=1}^{k}\gamma(k,k_n)\Big]
\delta_{m,\sum_{n\leq k}k_n}\delta_{k_k,k^\prime}}
{\sum_{k_1\ldots k_{k}}\Big[\prod_{n=1}^{k}\gamma(k,k_n)\Big]
\delta_{m,\sum_{n\leq k}k_n}}
\Big)
\nonumber
\\
&&\times
\Big(
\frac{\sum_{k_1\ldots k_{k^\prime}}
\Big[\prod_{n=1}^{k^\prime}\gamma(k^\prime,k_n)\Big]
\delta_{m^\prime,\sum_{n\leq k^\prime}k_n}\delta_{k_{k^\prime},k}}
{\sum_{k_1\ldots k_{k^\prime}}\Big[\prod_{n=1}^{k^\prime}\gamma(k^\prime,k_n)\Big]
\delta_{ m^\prime,\sum_{n\leq k^\prime}k_n}}
\Big)
\end{eqnarray}

\end{document}